\newcommand{\dR}{$\Delta R$~}
\newcommand{\ep}{$e^+$~}
\newcommand{\Erec}{$E_{rec}$}
\newcommand{\Ek}{$E_k$}

\newcommand{\rrr}{$r^3$}
\newcommand{\thetaaa}{$\theta $~}
\newcommand{\thetaPMT}{$\theta_{PMT}$}
\newcommand{\phii}{$\phi $~}
\newcommand{\Ri}{$\vec{R_i}$}
\newcommand{\rs}{$\vec{r}$}
\newcommand{\Ge}{$^{68}$Ge}
\newcommand{\Co}{${}^{60}$Co}
\newcommand{\Redep}{$\vec{r}_{edep}$}
\newcommand{\Rinit}{$\vec{r}_{init}$}

\documentclass[preprint,5p,number,times,sort&compress]{elsarticle}

\usepackage{lineno,hyperref}
\usepackage{color}
\usepackage{amsmath}

\modulolinenumbers[1]
\journal{Nuclear Instruments and Methods in Physics Research A}

\begin{document}

\begin{frontmatter}

\title{Improving the energy uniformity for large liquid scintillator detectors}

\author[firstaddress,secondaddress]{Guihong Huang}
\author[firstaddress]{Yifang Wang}
\cortext[mail]{Corresponding author}
\author[firstaddress]{Wuming Luo\corref{mail}}
\ead{luowm@ihep.ac.cn}
\author[firstaddress]{Liangjian Wen\corref{mail}}
\ead{wenlj@ihep.ac.cn}
\author[firstaddress]{Zeyuan Yu}
\author[firstaddress]{Weidong Li}
\author[firstaddress]{Guofu Cao}
\author[firstaddress]{Ziyan Deng}
\author[firstaddress]{Tao Lin}
\author[firstaddress]{Jiaheng Zou}
\author[thirdaddress]{Miao Yu}

\address[firstaddress]{Institute of High Energy Physics, Chinese Academy of Sciences, Beijing 100049, China}
\address[secondaddress]{School of Physical Sciences, University of Chinese Academy of Science, Beijing 100049, China}
\address[thirdaddress]{Wuhan University, Wuhan 430072, China}

\begin{abstract}
It is challenging to achieve high precision energy resolution for large liquid scintillator detectors.
Energy non-uniformity is one of the main obstacles. To surmount it, a calibration-data driven method was developed previously to
reconstruct event energy in the JUNO experiment. In this paper, 
we investigated the choice of calibration sources thoroughly, optimized the calibration positions
and corrected the residual detector azimuthal asymmetry. All these efforts lead to a reduction of the energy non-uniformity near the detector
boundary, from about 0.64\% to 0.38\%. And within the fiducial volume of the detector it is improved from 0.3\% to 0.17\%. As a result the energy resolution could be further improved.

\end{abstract}


\end{frontmatter}


\section{Introduction}
\noindent Liquid scintillator (LS) detectors with ultra-low background have been widely used in neutrino experiments. 
Just to name a few: KamLAND~\cite{kamland}, Borexino~\cite{borexino}, 
Double Chooz~\cite{doublechooz},  Daya Bay~\cite{dyb} and RENO~\cite{reno}.
Instead of dwelling on the outstanding scientific achievements made by these experiments 
in recent decades~\cite{kamland1, borexino1, doublechooz1, dyb1, reno1}, and how LS detectors will continue to play a crucial role in 
neutrino physics in the future, 
let us do a quick comparison of the LS detectors from above. The detector size varies from 10 tons to 10$^3$ tons in target mass,
and the detector energy resolution ranges from 5\% to 8\%. On the other hand JUNO~\cite{juno}
will be the largest LS detector in the world with a target mass of 20~kton upon completion, and its designed
energy resolution is $\sim$3\%/$\sqrt{E}$, which is much more precise with respect to earlier
experiments.
Even though it is rather challenging to achieve such high precision energy resolution for a large LS detector,
 a comprehensive calibration program~\cite{junocollaboration2020calibration} demonstrated 
that the required energy resolution of JUNO could be achieved, by accurately modeling 
the energy non-linearity and correcting for the energy non-uniformity.  The residual non-uniformity in~\cite{junocollaboration2020calibration}
is less than 0.3\% within the fiducial volume of the detector. Since the energy resolution has such significant impact, 
and  one of the main contributing factors is the energy non-uniformity, we wanted to further reduce it, especially for regions near the detector boundary  which amounts to more than 20\% of the whole detector volume.

Due to the complicated optical processes, an optical model independent method~\cite{Wu_2019} 
was developed previously to reconstruct the energy of positrons from inverse $\beta$-decay (IBD) 
events in the Central Detector (CD) of JUNO.  The basic idea is to construct the maps of expected 
photoelectrons (PEs) for Photomultiplier tubes (PMTs)  from  calibration 
data, and then use these maps to build a maximum likelihood function to reconstruct the event energy.
In this paper we will further improve the energy uniformity in the CD of JUNO, by taking into account the asymmetry of a 
realistic detector and optimizing the calibration strategy. Due to lack of real data, 
Monte Carlo (MC) simulation data generated by JUNO offline software~\cite{sniper} are used instead. 
The ideas and methods discussed here are also applicable to other experiments using large LS detectors.

The structure of this paper is as follows: in Sec.~\ref{sec:detector}, we will briefly describe 
the CD of JUNO and its calibration systems. All the MC samples used will be listed 
in Sec.~\ref{sec:samples}. From Sec.~\ref{sec:ErecMu} to Sec.~\ref{sec:position}, we will present an update on the
maps of expected PEs, a thorough study on the calibration sources, and an optimization on calibration 
points respectively. In Sec.~\ref{sec:phicorr} we will discuss the residual energy non-uniformity.
And finally we will give the summary in Sec.~\ref{sec:summary}.

\section{JUNO CD and Calibration Systems}
\label{sec:detector}
\begin{figure}[!ht]
\centering
\includegraphics[width=0.4\textwidth,,angle=0]{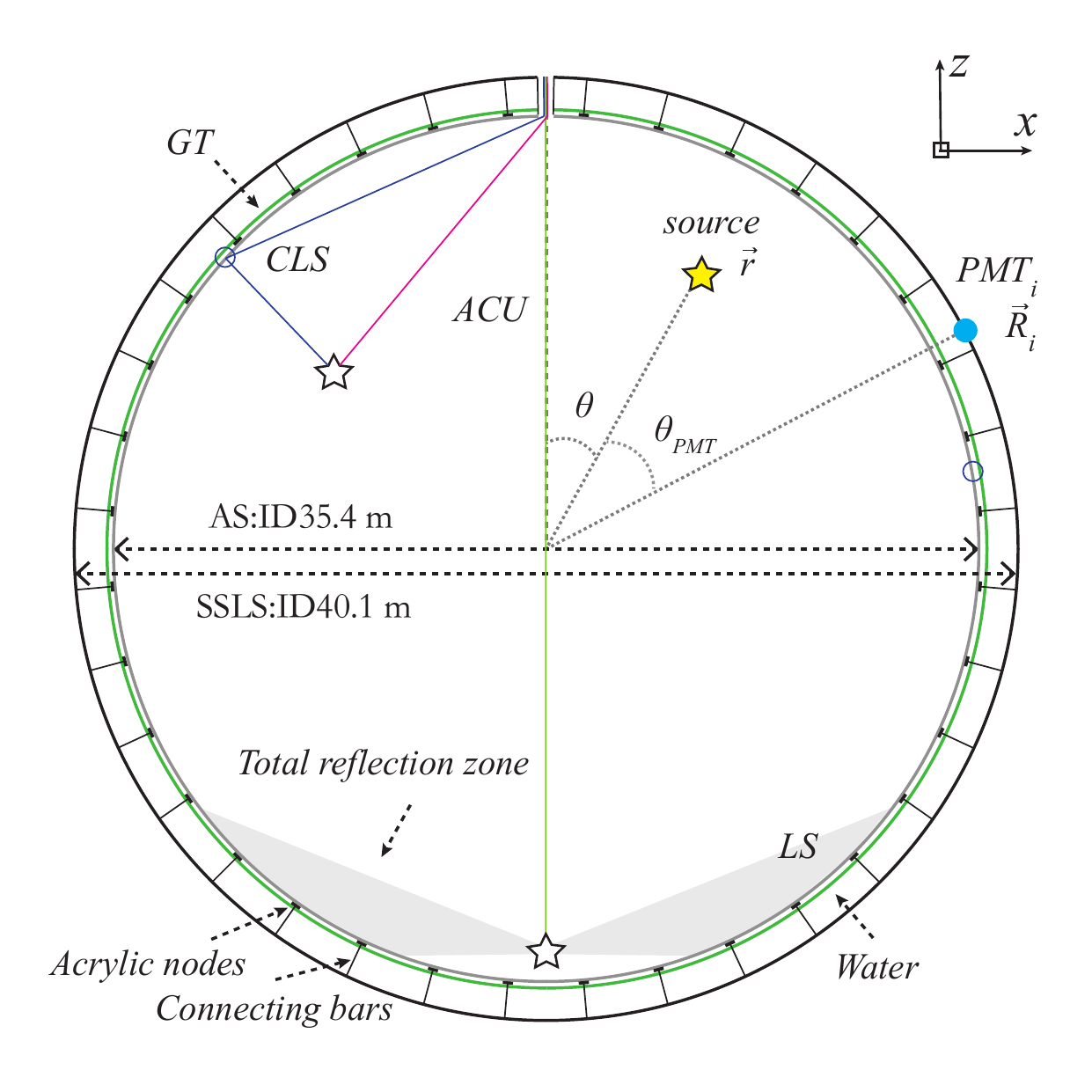}
\caption{ Schematic view of the CD and the calibration system. Z-axis is the vertical central axis of the CD. An example of total reflection is also shown at the bottom. Upper-Right: definition of the three parameters of $\hat{\mu}$ in Sec.~\ref{sec:ErecMu}. \rs ~is the calibration source position, 
\Ri ~is the $i^{th}$ PMT position and $\theta_{PMT}$ is the angle between \rs ~and \Ri. }
\label{fig:MapPara}
\end{figure}

 \noindent A schematic view of the CD and the calibration system of JUNO is shown in Fig.~\ref{fig:MapPara}. The CD is made up of an acrylic sphere which has a diameter of 35.4~m and contains 20,000 ton LS. The acrylic sphere is supported by a stainless-steel latticed shell (SSLS) via acrylic nodes and connecting bars. The diameter of the SSLS is 40.1~m, and the gap between it and the acrylic sphere is filled with pure water. About 17,600 20-inch PMTs and 25,600 3-inch PMTs are installed on the stainless- steel shell to collect photons. Given the different refractive indices of LS and water, refraction and total reflection could occur during photon propagation.
 JUNO also has a complex calibration system~\cite{junocollaboration2020calibration} which consists of four sub-systems, namely the Automated Calibration Unit (ACU)
, the Cable Loop System (CLS), the Guide Tube (GT) and the remotely operated vehicle (ROV). Only the former three are used for
energy reconstruction in this paper. It should be emphasized that each
sub-system could cover a different detector region: ACU can move along the Z-axis of the CD, CLS is able to reach those points  
 permitted by the mechanics of the loop system within X-Z plane.  GT is mounted 
on the outer surface of the acrylic sphere, designed to calibrate the detector in the edge region complementary
to ACU and CLS. 

\section{Monte Carlo Samples}
\label{sec:samples}
\noindent Various calibration data samples with different sources taken from Ref.~\cite{junocollaboration2020calibration} are produced. For the prompt signal of IBD events,
a set of positron samples are also prepared.
The information of the calibration samples and the positron samples used in 
this paper are summarized in Tab.~\ref{tab:CalibSampleInfo} and Tab.~\ref{tab:PhySampleInfo}, respectively.
For all these samples,  the detector simulation is done based on Geant4~\cite{G4}. LS properties~\cite{LS1} and optical processes
of photons propagating in LS are implemented~\cite{LS2, LS3}. Realistic detector geometry such as the arrangement 
of the PMTs and the supporting structures is also deployed.  For simplicity the electronics simulation which includes various PMT characteristics is disabled.

The calibration samples are used to construct the maps of expected PEs per unit energy for PMTs, 
referred to as $\hat{\mu}$ hereafter and described in detail in Sec.~\ref{sec:ErecMu}. Calibration sources
with different types and energies are compared in order to select the most suitable one. Nine sets of positron samples with 
kinetic energy \Ek~= (0, 1, 2, ..., 8)~MeV are used to evaluate the performance of energy reconstruction. The events in each 
 positron sample are uniformly distributed in the CD.

\begin{table}[h]
\centering
\caption{Information of the calibration samples. $^{68}$Ge is a positron emitter, the kinetic energy of the positrons will be absorbed by the source enclosure, so only the annihilation gammas are released. For the Laser source, ``op" stands for optical photon and 1~MeV corresponds 
to  11522 optical photons. The event statistics per position is 10k. }
\begin{tabular}{ccccc}
\hline
\hline
    Source& Type& Energy [MeV]& N$_{position}$ & Stats./pos.\\
\hline
    \Ge& $\gamma$& 2$\times$0.511& 2000 & 10k\\
    \Co& $\gamma$& 1.173 + 1.333& 275 & 10k\\
AmC& (n,H)$\gamma$& 2.22& 275 & 10k\\
    Laser& op&  1& 2000 & 10k\\
\hline
\end{tabular}
\label{tab:CalibSampleInfo}
\end{table}

\begin{table}[h]
\centering
\caption{List of the positron samples.}
    \begin{tabular}{cccc}
\hline
\hline
        Source& Kinetic energy& Statistics/MeV& Position\\
\hline
    $e^+$& (0,1,2, ..., 8)~MeV& 450k & uniform in CD\\
\hline
\end{tabular}
\label{tab:PhySampleInfo}
\end{table}

\section{Energy Reconstruction  and  $\hat{\mu}$}
\label{sec:ErecMu}
\noindent As described in Ref.~\cite{Wu_2019}, an optical model independent method
was developed to reconstruct the energy of positrons in the JUNO CD. The observables for each positron
are \{$k_{i}$\}, where $k_{i}$ represents the number of detected PEs for 
the $i^{th}$ PMT and is expected to follow a Poisson distribution. 
The mean value of the Poisson distribution $\mu_{i}$ is the product of the positron visible energy 
$E_{vis}$ and $\hat{\mu_{i}}$ from Sec.~\ref{sec:samples}.
So the probability of observing \{$k_{i}$\} for all PMTs can be constructed as Eqn.~\ref{eq:PEMLE_LF}  
when an event deposits energy at position ($r, \theta, \phi$). 

\begin{equation}
  \begin{gathered}
\mathcal{L}( \{k_{i}\}|r,\theta,\phi,E_{vis})  = 
    \prod_i \mathcal{L}(k_{i}|r,\theta,\phi,E_{vis})=\prod_i \frac{e^{-\mu_{i}}\cdot \mu_{i}^{k_{i}}}{k_{i}\!} \\
    \mu_{i} = E_{vis} \cdot \hat{\mu_{i}}
      \end{gathered}
\label{eq:PEMLE_LF}
\end{equation}

\noindent where the index $i$ runs over all PMTs. 
After obtaining $\hat{\mu_{i}}$, the event energy can be fitted by maximizing 
this likelihood function.
In order to decouple the influence of the vertex uncertainty on the energy reconstruction, the event vertex 
 is assumed to be known in this study.

The key component of the energy reconstruction method discussed above is $\hat{\mu}$.
In Ref.~\cite{Wu_2019}, it is derived from the ACU calibration data, under the assumption
that the JUNO CD has good spherical symmetry. If the calibration source position is defined
as \rs~= ($r, \theta, \phi$ = 0) and the $i^{th}$ PMT position as \Ri, as shown in Fig.~\ref{fig:MapPara}, 
then $\hat{\mu}$  can be calculated as: 

\begin{equation}
\begin{gathered}
    \hat{\mu}(r, \theta_{PMT}) =  \frac{\mu(r, \theta_{PMT})}{ E_{vis} }= ( \frac{1}{M} \sum_{i=1}^{M}  \frac{\bar{n}_{i}} {DE_{i}}  ) \cdot \frac{1}{E_{vis}}\\
    E_{vis} = PE_{total} / Y_0
\end{gathered}
\label{eq:EqMu0}
\end{equation}
where $E_{vis}$ is the visible energy of the calibration source,
$PE_{total}$ is the total number of PEs,  $Y_0$ is the constant light yield defined in Ref.~\cite{junocollaboration2020calibration},
the index $i$ runs over the PMTs with the same $\theta_{PMT}$, 
$\bar{n}_{i}$ is the average number of detected PEs and $DE_{i}$ is the relative detection 
efficiency. Given there are only finite ACU calibration positions,  $\hat{\mu}(r = z, \theta_{PMT})$
from these positions are extrapolated through linear interpolation to the entire $(r, \theta_{PMT})$ phase space.

\begin{figure}[!ht]
\centering
\includegraphics[width=0.42\textwidth,,angle=0]{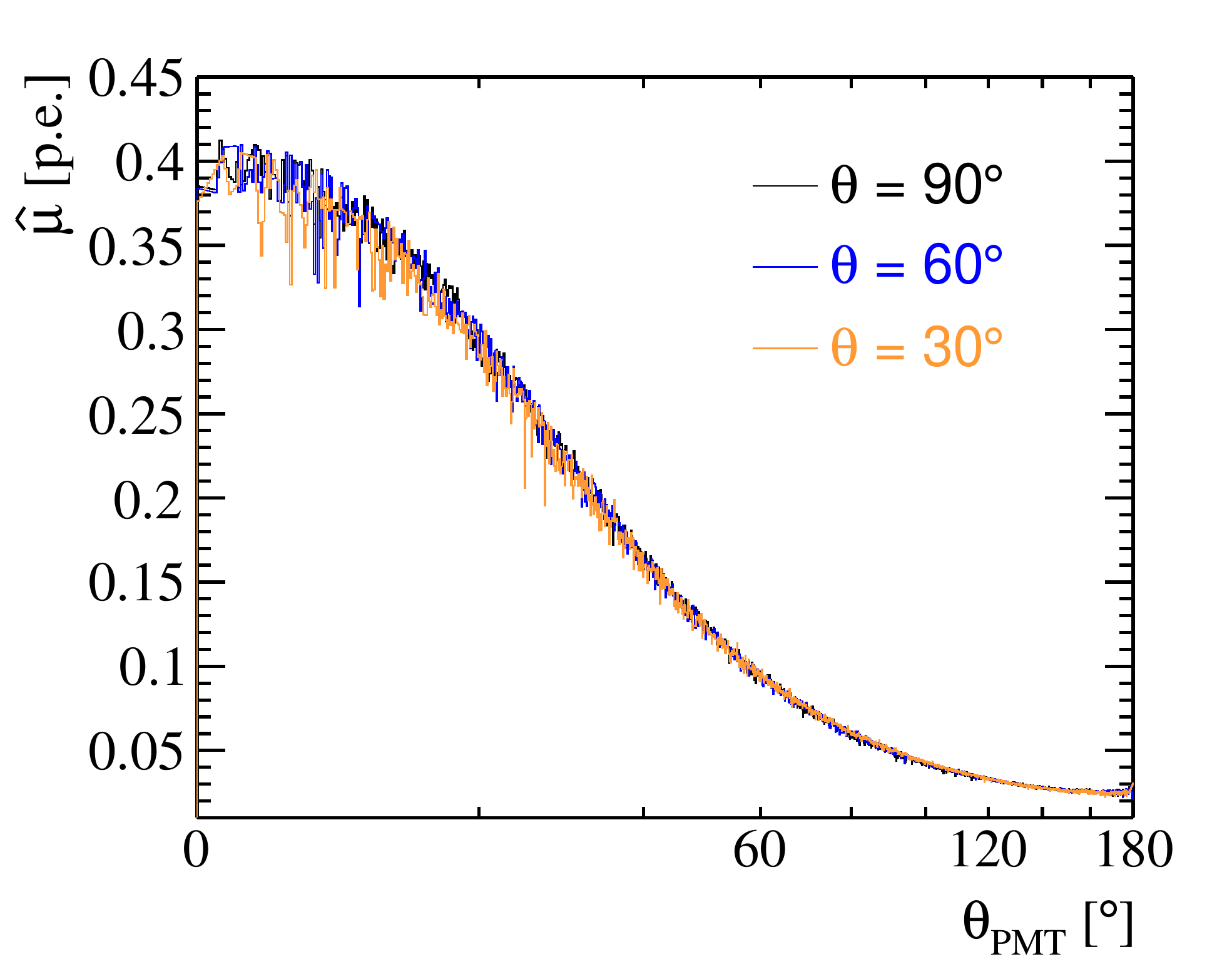}
    \caption{Comparison of $\hat{\mu}$ for calibration positions with different  $\theta$ angle and fixed radius $r$ = 10~m.
    The spikes are mainly caused by the shadowing effect of the the acrylic nodes and connecting bars. The unit of $\hat{\mu}$ is p.e. which stands for 1 photon electron.}
\label{fig:nPEResp}
\end{figure}

Fig.~\ref{fig:nPEResp} compares the $\hat{\mu}(r, \theta_{PMT})$ maps
for calibration positions with the same radius but different $\theta$ angle, 
as could be collected by the CLS calibration sub-system.
The apparent differences, which are mainly caused by the shadowing effect of the acrylic nodes and connecting bars when $\theta$ varies, indicate that the detector is not symmetric along the \thetaaa direction,
and this \thetaaa dependence for $\hat{\mu}(r, \theta_{PMT})$ must be taken into account. 
Since the CLS system can move in the X-Z plane,  we could combine the CLS and ACU calibration data 
to construct $\hat{\mu}(r, \theta, \theta_{PMT})$ in the same way as before.

\begin{figure}[!ht]
\centering
  \includegraphics[width=0.32\textwidth,angle=0]{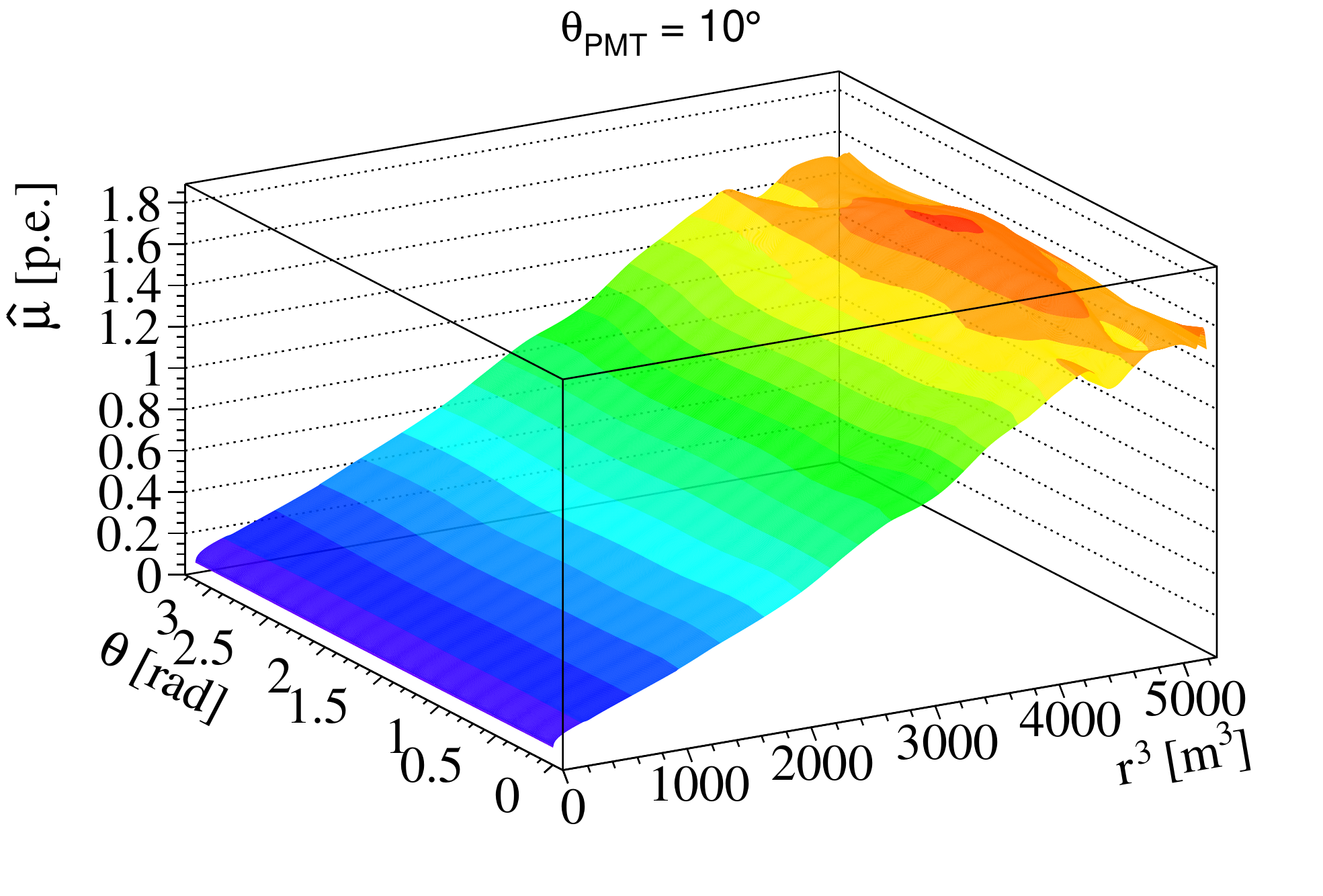}
  \includegraphics[width=0.32\textwidth,angle=0]{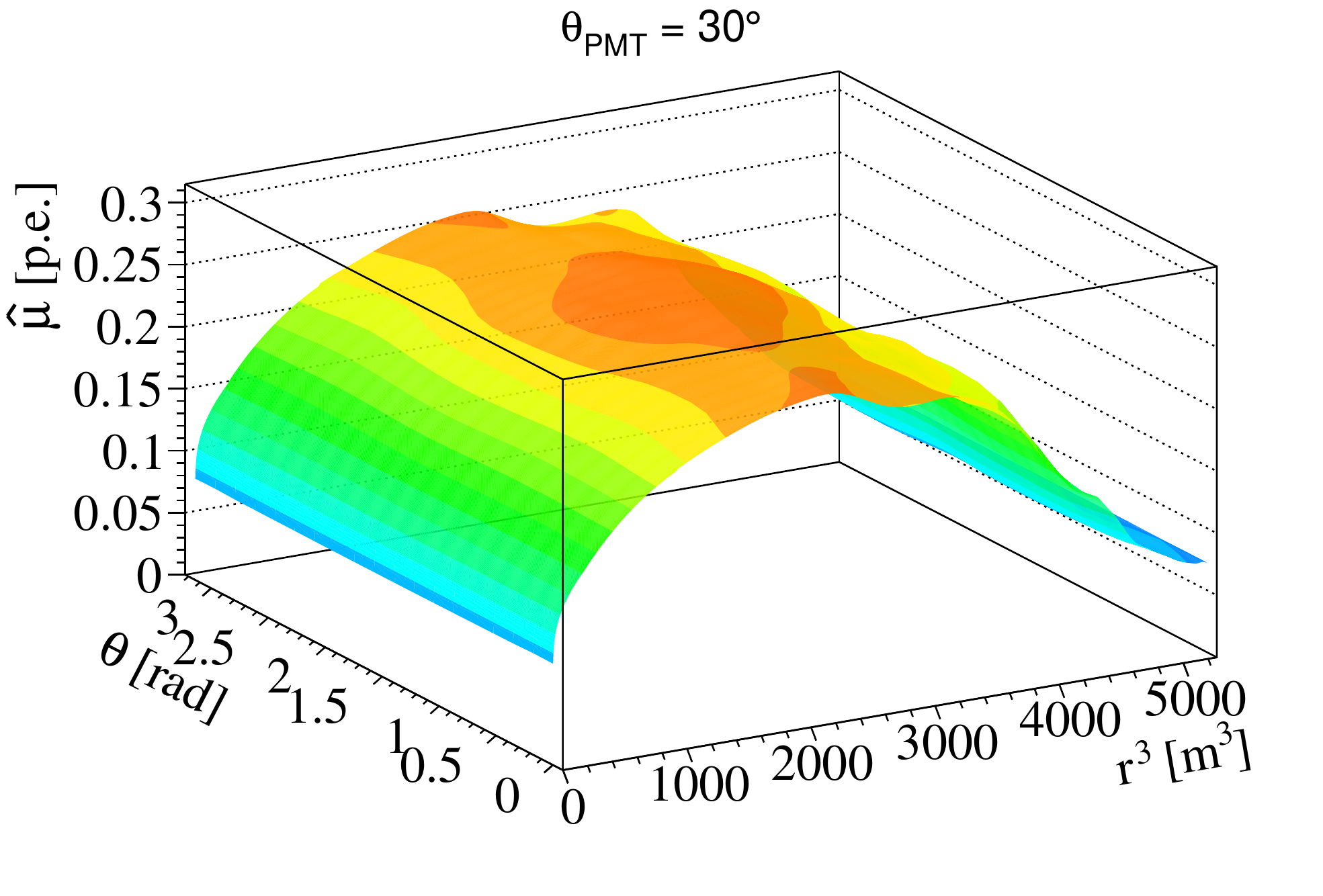}
  \includegraphics[width=0.32\textwidth,angle=0]{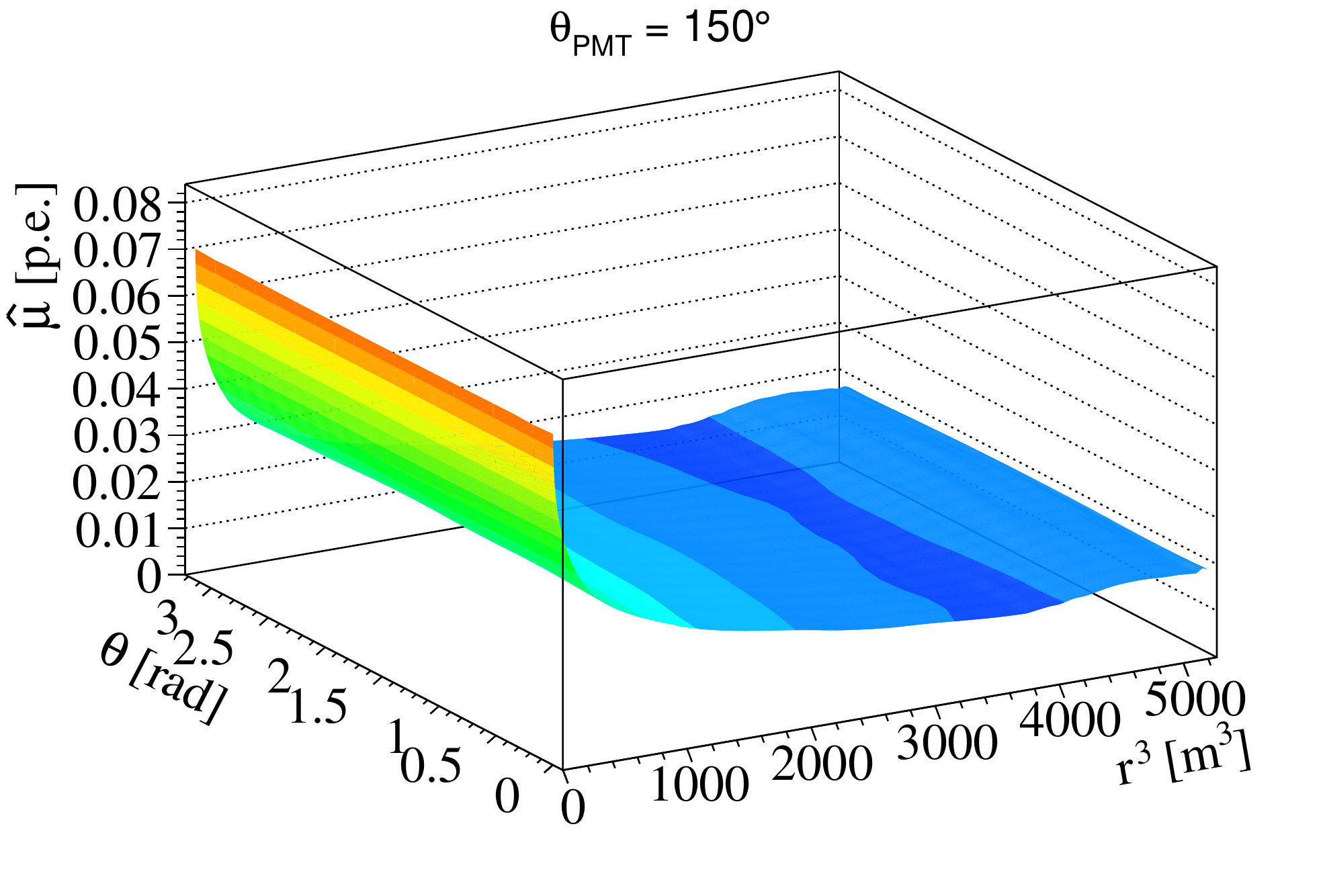}
\caption{Examples of the $\hat{\mu}(r, \theta, \theta_{PMT})$ maps at three $\theta_{PMT}$ angles.}
\label{fig:nPEMap}
\end{figure}

A few examples of the $\hat{\mu}(r, \theta, \theta_{PMT})$ maps at fixed $\theta_{PMT}$ values are shown in Fig.~\ref{fig:nPEMap}. 
 And they are also affected by the same shadowing effect.
 The Delaunay triangles based cubic spline interpolation has been applied to  $\hat{\mu}(r, \theta, \theta_{PMT})$,
so that it could be extrapolated to the whole ($r, \theta$) phase space from finite calibration positions.
At this point, it is quite natural to ask whether there is any \phii dependence for $\hat{\mu}$, which 
could be caused by any detector asymmetry along the \phii  direction. We will leave this discussion 
to Sec.~\ref{sec:phicorr}.

\section{Comparison of Calibration Sources}
\label{sec:source}
\noindent Our energy reconstruction method heavily relies on the usage of calibration data. 
Given all the available calibration sources, which one gives the best energy reconstruction performance?
Bearing this question in mind, we thoroughly investigated these sources: other than the energy, 
what else could be different for these sources? How do the $\hat{\mu}$ maps compare? And eventually
how does the energy reconstruction performance compare?

	\begin{figure}[!ht]
		\centering
		\includegraphics[width=0.42\textwidth]{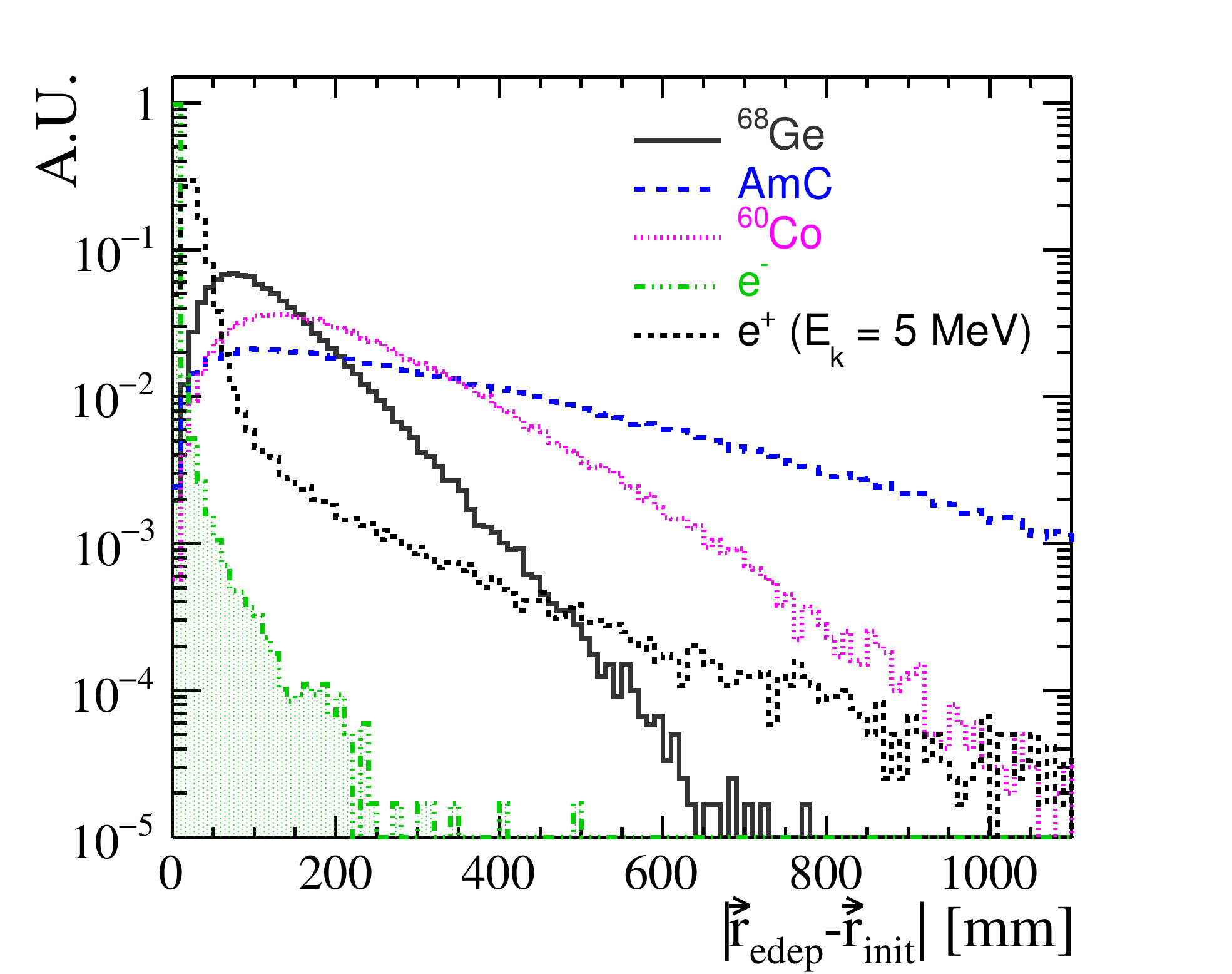}{\centering}		
		\caption{Distribution of the distance \dR between energy-deposit center \Redep ~and initial calibration position
\Rinit ~for different sources. A hypothetical position source with $Ek$ = 5~MeV is also drawn for comparison.}
		\label{fig:vtxSmear}			
	\end{figure}   	

	\begin{figure*}[ht!]
		\centering
		\includegraphics[width=0.96\textwidth]{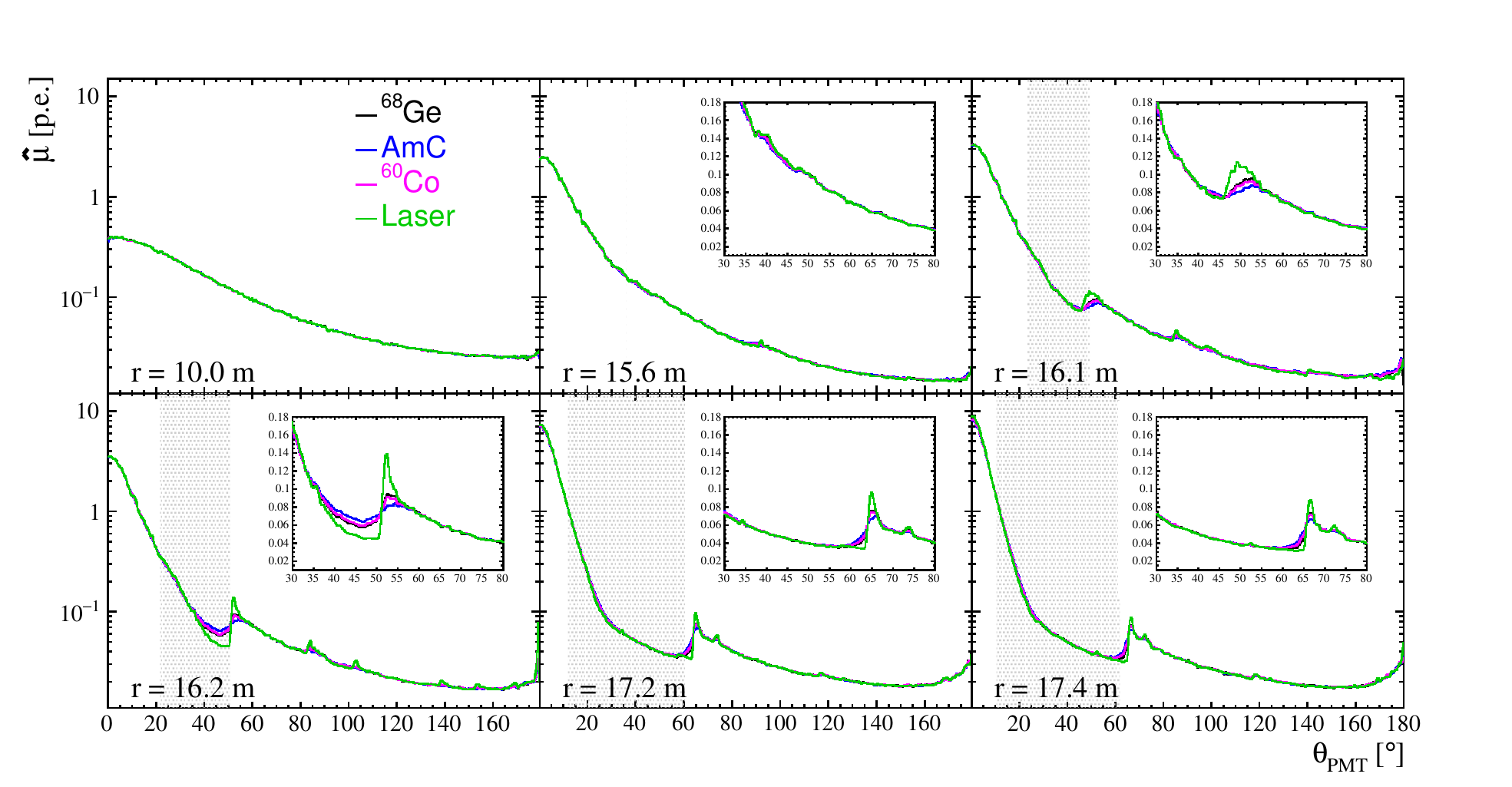}{\centering}		
		\caption{Comparison of  $\hat{\mu}$ for different sources at a few representative calibration positions.
From left to right and top to bottom, $r$ = (10, 15.6, 16.1, 16.2, 17.2, 17.4)~m, \thetaaa= 90$^\circ$, \phii= 0$^\circ$. 
The shaded region indicates the total reflection zone.}
		\label{fig:mapCompSource}			
	\end{figure*}
	
In Fig.~\ref{fig:vtxSmear} we compared the distribution of the distance \dR between energy-deposit 
center \Redep ~and initial calibration position \Rinit ~for different sources. Laser source is set to be point-like in 
the MC simulation, and in reality it is approximately point-like due to the diffuse ball which absorbs the optical photons from Laser and re-emits them isotropically~\cite{CalibLaser}.  For illustration purpose, a virtual mono-energetic electron source is depicted instead of Laser.
A hypothetical position source with $E_k$ = 5~MeV is also drawn for comparison. Electron source is very close to point-like and \Redep ~is almost identical to \Rinit. While for the gamma sources 
 there is a clear deviation of \Redep ~from \Rinit, since a gamma 
deposits its energy in LS mainly through multiple Compton scattering. For \Ge~source, the two
 gammas from positron-electron annihilation tend to be back to back directionally. 
On the other hand, \Co~radiates two gammas (1.173~MeV and 1.333~MeV) without any 
direction correlation, leading to a wider spread of $\Delta R$. AmC is a neutron source, 
the neutron will travel some distance before being captured by hydrogen and then emitting 
a 2.22~MeV photon, thus its \dR~has the widest spread among all the sources.

The CD is divided into regions I, II and III: namely the central region ($r<$ 15.6~m), the total reflection region 
(15.6~m $ < r <$ 17.2~m) and the outer-FV (fiducial volume) region (17.2~m $< r$), and six representative calibration positions are picked: 
 $r$ = (10, 15.6, 16.1, 16.2, 17.2, 17.4)~m, \thetaaa= 90$^\circ$, \phii= 0$^\circ$. The comparison of the $\hat{\mu}$ maps among different 
sources at these points is shown in Fig.~\ref{fig:mapCompSource}. The $\hat{\mu}$ maps had been smoothened and it was checked that this
smoothening has negligible impact on the energy reconstruction.
A few observations immediately stand out:
\begin{itemize}
\item[1.] In region I, the $\hat{\mu}$ maps are nearly monotonic and also similar for all sources.
\item[2.] In regions II and III, the $\hat{\mu}$ maps have a kink. And there are noticeable discrepancies 
 among the sources around the kink.
 \end{itemize}

\noindent In region I where the sources 
are relatively far away from the PMTs, all sources could be approximately regarded as point-like,
thus leading to similar $\hat{\mu}$ maps. Moreover, since total reflection won't occur in region I, as \thetaPMT 
~increases, smaller solid angle leads to decreased $\hat{\mu}$. While in regions II and III, there is always a total 
reflection zone for any given source position, mainly due to the refractive index mis-match between LS (n = 1.496@430~nm) and water (n = 1.353@430~nm). An example is shown in Fig.~\ref{fig:MapPara}. For those PMTs in the total reflection zone
(as indicated by the shadowed region in the plots), 
one would expect a large decrease of detected PEs due to photon redistribution  and loss. 
In addition \dR also becomes relevant for these PMTs, because any small deviation of the source 
position would partially mitigate the impact of total reflection. And the more spread \dR is, the 
stronger the mitigation effect is, which is  illustrated by the enlarged figures in Fig.~\ref{fig:mapCompSource} 
and Fig.~\ref{fig:vtxSmear} from above.

\begin{figure}[!ht]
		\centering
		\includegraphics[width=0.42\textwidth]{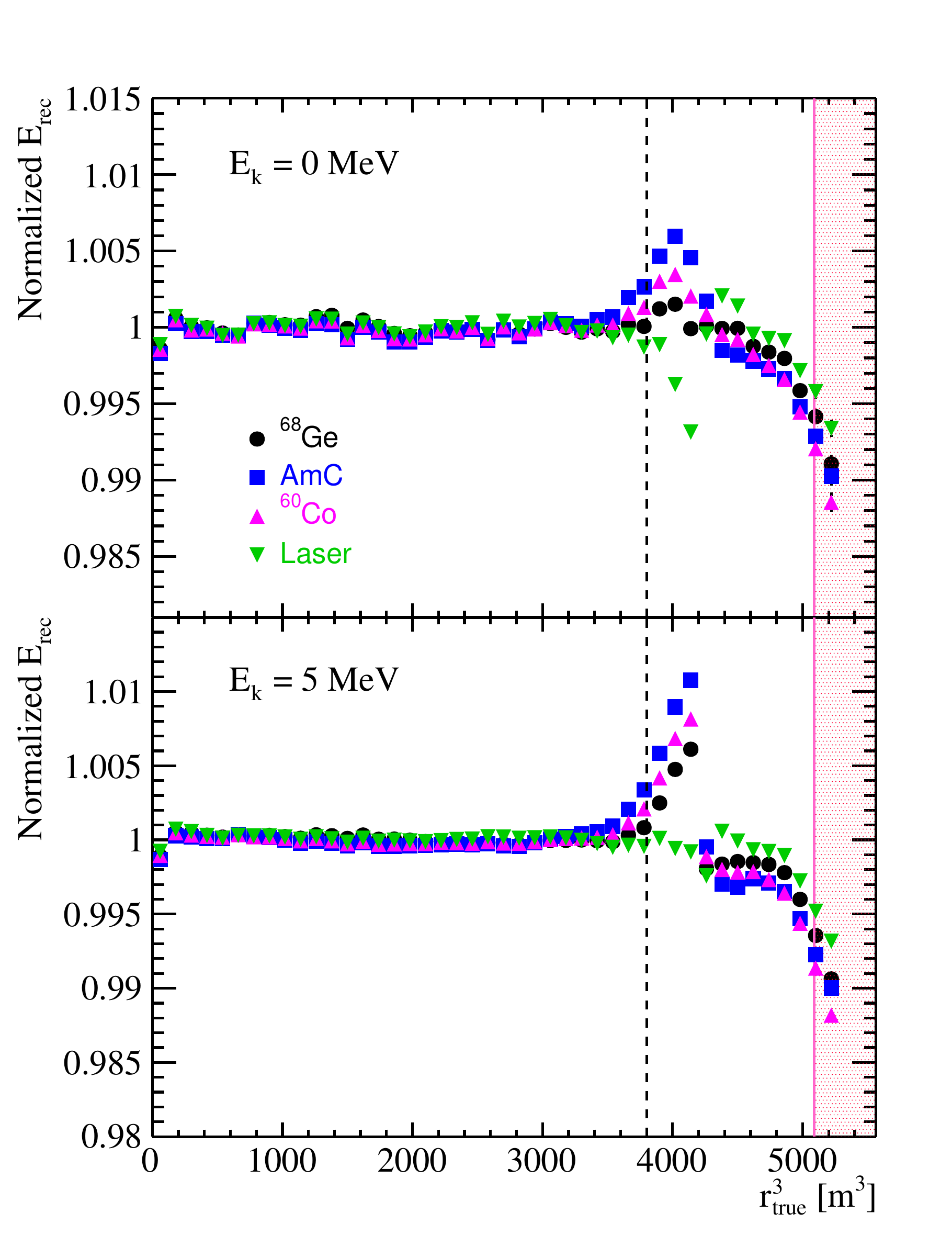}{\centering}
		\caption{Uniformity of reconstructed energy \Erec~with respect to \rrr~using $\hat{\mu}$ maps from different sources. 
		For each \rrr~bin, the mean value of \Erec~is plotted. Top and bottom plots correspond to \Ek~= 0, 5~MeV \ep samples respectively.} 
		\label{fig:nonUniCompSource}			
	\end{figure}

After obtaining the $\hat{\mu}$ maps using different sources and the calibration points from Case 2 in Sec.~\ref{sec:position}
, we applied them individually to the energy 
reconstruction of positron samples listed in Tab.~\ref{tab:PhySampleInfo}. The  
uniformity of the reconstructed energy \Erec ~with respect to \rrr ~for two different energies 
was plotted in Fig.~\ref{fig:nonUniCompSource}.
The two vertical lines indicate the boundaries of the three regions. Each curve is normalized by 
its average value within region I. 
The results of \Erec~are quite consistent among the sources in region I for all positron energies, which is not surprising given that 
the $\hat{\mu}$  maps from different sources are almost the same. The non-uniformity in region II could be traced back 
to the features of the $\hat{\mu}$  maps, caused by total reflection as mentioned before. Take the bump peak in 
the \Ek~= 5~MeV case as an example. The corresponding radius is $r$ = 16.1~m which is the same as the top right plot 
in Fig.~\ref{fig:mapCompSource}. Comparing to the Laser source, the other sources will give smaller 
expected $\hat{\mu}$, resulting in larger \Erec. The size of the non-uniformity for each source is positively 
correlated to its \dR spread.
Another important thing we should note is that using the $\hat{\mu}$ maps from \Ge~source yields the best 
uniformity at \Ek~= 0~MeV, while at high energies the $\hat{\mu}$ maps from Laser source perform the best. 

By comparing the sources thoroughly, we aimed to pick out one that gives good energy reconstruction performance 
across the entire positron energy range. Based on the studies above, none of the sources is satisfactory. 
If one single source won't do, is it possible to use a combined source? Let us dive back to the energy
deposition of positron in LS again. The whole process can be naturally broken down  into two parts:  
almost all positrons will fully deposit their kinetic energy first, this part can be treated as a point-like source. 
There is a small probability that positrons will annihilate in flight, but this can be safely ignored.
The second part is the positron-electron annihilation producing two gammas, which is almost the same 
as the \Ge~source. This explains why the \Ge~source performs the best for \Ek~= 0~MeV positron
events. With increasing kinetic energy, positron becomes more and more point-like. Consequently point-like 
source such as Laser is more suitable at higher energies. Thus for positrons with visible energy $E_{vis}$, we propose the 
following combined $\hat{\mu}^{comb}(r,\theta, \theta_{PMT})$:

	\begin{equation}
	\begin{gathered}
		\hat{\mu}^{comb} =\frac{1}{E_{vis}} \cdot ( E^{Ge}_{vis} \cdot \hat{\mu}^{Ge}(r,\theta, \theta_{PMT}) + E_k \cdot \hat{\mu}^{L}(r,\theta, \theta_{PMT}) )\\
		E_{vis} = E^{Ge}_{vis} + E_k 
		 \label{equ:expnpe}
 	\end{gathered}
	\end{equation}	
where $\hat{\mu}^{Ge}(r, \theta, \theta_{PMT})$ and $\hat{\mu}^{L}(r, \theta, \theta_{PMT})$ correspond to the 
annihilation part and kinetic energy part of positron respectively, $E_k$ is the kinetic energy of positron and $E^{Ge}_{vis}$ (1.022~MeV) is the visible energy of $^{68}$Ge.

	\begin{figure}[!ht]
		\centering
		\includegraphics[width=0.42\textwidth]{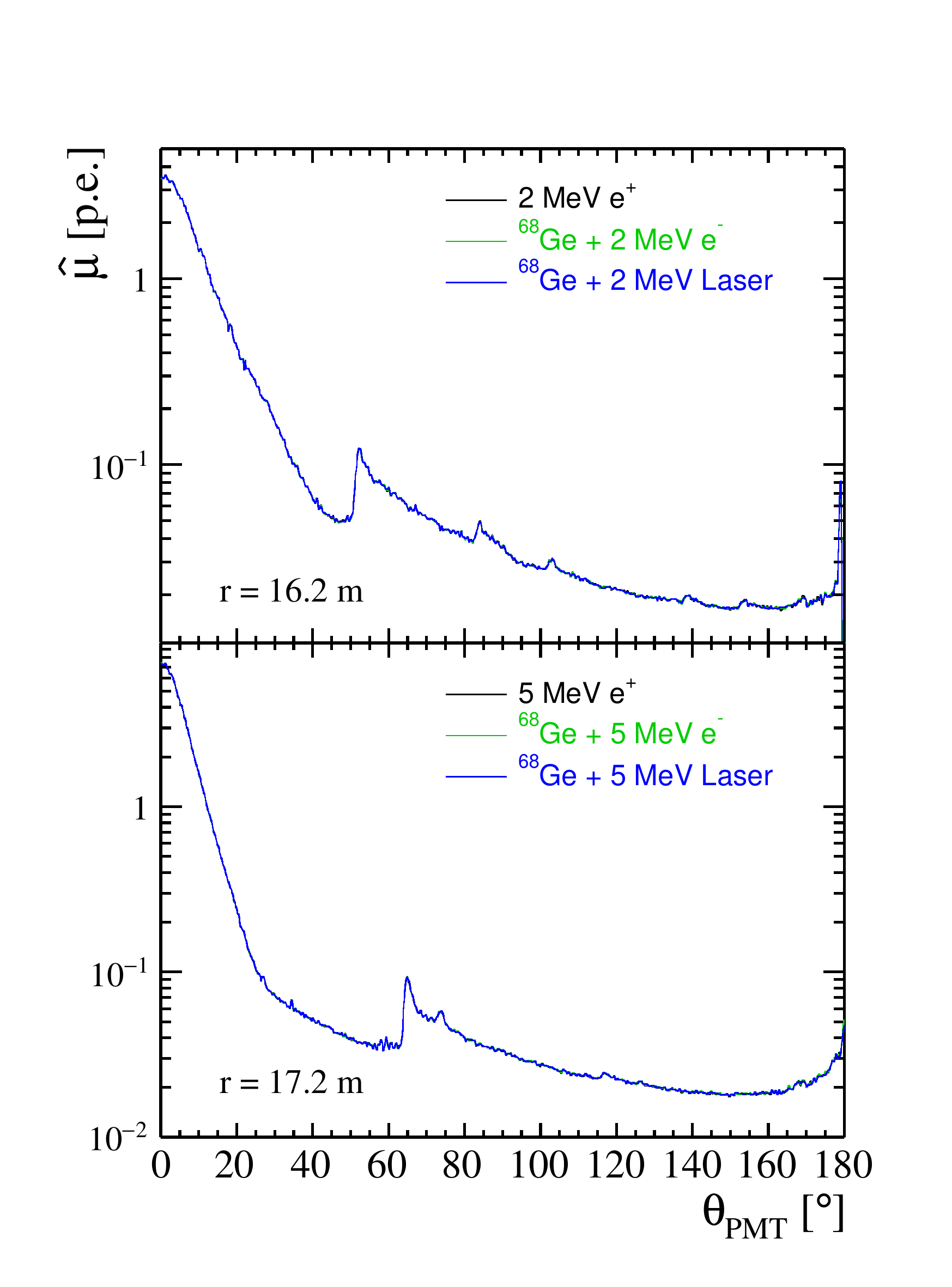}{\centering}		
		\caption{Validation of the combined maps $\hat{\mu}^{comb}$. Top and bottom plots are for \Ek~= 2, 5~MeV and $r$ = 16.2, 17.2~m respectively. The three curves in each plot overlap with each other, confirming the correctness of the combined maps.} 
		\label{fig:combinedMap}			
	\end{figure}   	

	\begin{figure*}[!ht]
		\centering
		\includegraphics[width=0.95\textwidth]{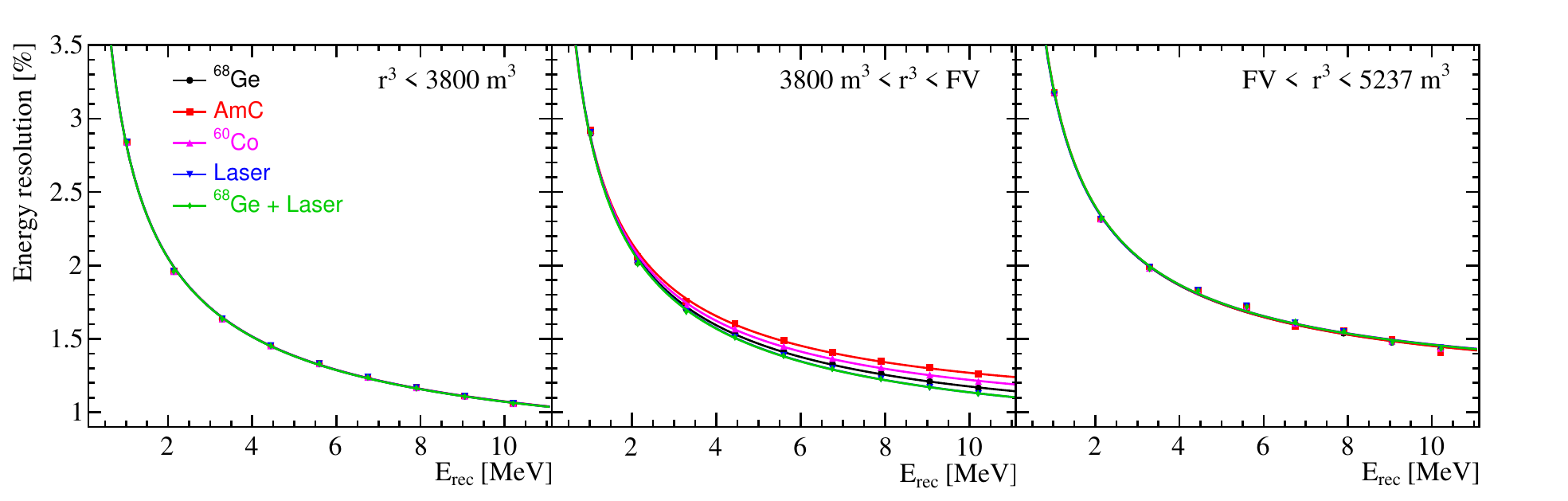}{\centering}		
		\caption{Comparison of the resolution of the reconstructed visible energy for positrons in the three regions of the CD. Different colors correspond to the various sources which are used to construct the $\hat{\mu}$ maps. The dots correspond to positrons with different kinetic energies. \Erec~is the mean value of the reconstructed visible energy. Please note that energy non-linearity is not corrected here. As one can see from the middle plot, with better energy uniformity, the combined source is able to improve the energy resolution in Region II.}
		\label{fig:Eres1}			
	\end{figure*}   	
	
To validate the combined maps $\hat{\mu}^{comb}$, they were compared to those produced with positron samples listed in Tab.~\ref{tab:PhySampleInfo}. Across the whole energy range, $\hat{\mu}^{comb}$ are able to match the  positron $\hat{\mu}$ maps. A few examples are shown in Fig.~\ref{fig:combinedMap}.
Note that it is assumed the kinetic energy part of the combined maps is linearly proportional to the kinetic energy.  
Energy non-linearity is not considered and has tiny impact on $\hat{\mu}^{comb}$. Replacing Laser with other point-like 
sources such as electron works as well and does not make any big difference for $\hat{\mu}^{comb}$.   
Laser is chosen but not electron simply due to the lack of mono-energetic electron sources in reality.
For the Laser source, the emitted photons are assumed to be isotropic. 
In reality the non-uniformity of photon emission for the Laser source in JUNO is expected to be about 
a few percent. Alternative Laser samples were produced where an arbitrary 5\% non-uniformity 
is added by hand. The resulting $\hat{\mu}^L$ maps do not change much compared with the 
default Laser samples, indicating that our calibration procedure is not particularly sensitive to the non-uniformity of the Laser source.

\begin{figure}[!ht]
		\centering
		\includegraphics[width=0.42\textwidth]{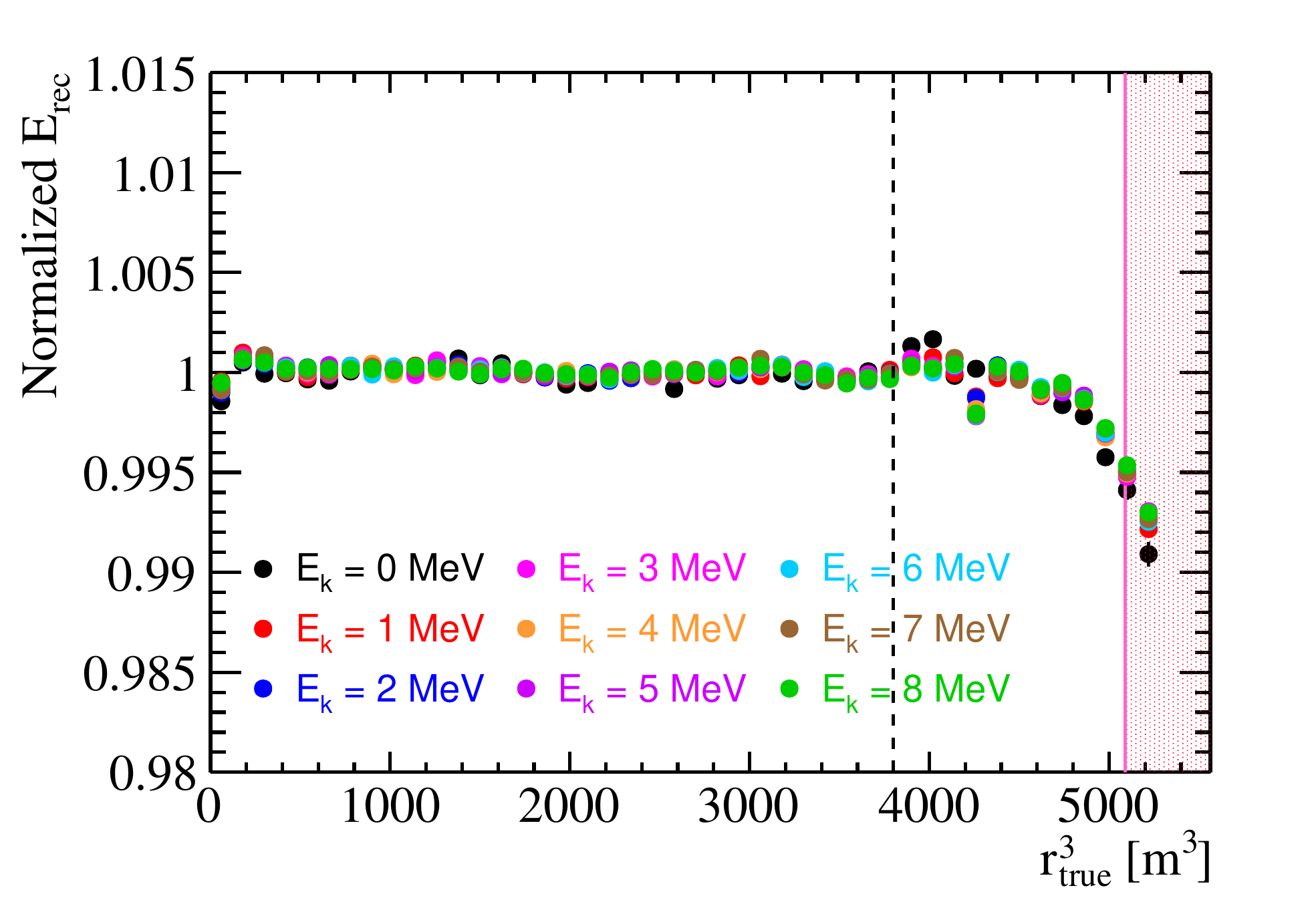}{\centering}		
		\caption{Uniformity of \Erec ~with respect to \rrr ~using combined maps $\hat{\mu}^{comb}$ at various energies.
		For each \rrr~bin, the mean value of \Erec~is plotted.}
		\label{fig:nonUniNew}			
	\end{figure}

After $\hat{\mu}^{comb}$ were produced and validated, their energy reconstruction performance was 
evaluated as before. The energy uniformity at various energies is shown in Fig.~\ref{fig:nonUniNew}, 
 which clearly has very small dependence on the energy. More importantly, it is largely improved in the total reflection region
 comparing to Fig.~\ref{fig:nonUniCompSource}. 	
For each positron sample with different kinetic energy, the distribution of the reconstructed visible energy is fitted with a Gaussian
function, and the corresponding energy resolution is defined as the ratio of the Gaussian sigma to the Gaussian mean.
The energy resolution with respect to the mean value of the reconstructed visible energy for all the positron samples in the three regions is shown in Fig.~\ref{fig:Eres1}. Different colors correspond to different sources that are used to produce the $\hat{\mu}$ maps.
The dots represent the energy resolution at different energies. Please note that energy non-linearity is not corrected here.
In regions I and III, the energy resolution is almost identical when using $\hat{\mu}$ maps from different sources.
In region II, the combined maps $\hat{\mu}^{comb}$ yield the best energy resolution especially at high energies, which is 
a direct consequence of improved energy uniformity. 
The energy resolution is fitted with an empirical model below:

	\begin{equation}
	\label{eqn:EresFit}
		\frac{\sigma}{E} =\sqrt{\frac{a^2}{E} + b^2}
	\end{equation}		

\noindent where $a$ is related to the photon poisson statistics, $b$ is the constant term and the unit of energy is MeV. The fitted results are summarized in Tab.~\ref{tab:fittedResults}.
\begin{table}[h!]
  \begin{center}
    \caption{Comparison of fitted parameters for the empirical energy resolution model among various sources. The unit for a (b) is 
    \%$\times$MeV$^{\frac{1}{2}}$ (\%). The fit uncertainties for a and b are less than 0.005 (0.02) in regions I and II (III).}
    \label{tab:fittedResults}
    \begin{tabular}{ccccccc}
      \hline
      \hline
      \textbf{Region} & \multicolumn{2}{c}{\textbf{I}} & \multicolumn{2}{c}{\textbf{II}} & \multicolumn{2}{c}{\textbf{III}}\\
      \hline
         & a & b & a & b & a & b \\

\hline
        AmC&                                2.76 & 0.626 & 2.74 & 0.926 & 3.02 & 1.09 \\
        \Co&                       2.76 & 0.624 & 2.76 & 0.853 & 3.01 & 1.11 \\
        \Ge&                       2.76 & 0.623 & 2.77 & 0.784 & 3.01 & 1.10 \\
        Laser&                               2.76 & 0.623 & 2.80 & 0.711 & 3.00 & 1.11 \\
        \Ge+Laser&            2.76 & 0.622 & 2.79 & 0.715 & 3.01 & 1.11 \\
      \hline
      \hline
    \end{tabular}
  \end{center}
\end{table}
 In regions I and III, there is no big difference for $a$ and $b$. But in region II, it is clear that energy non-uniformity 
contributes to the $b$ term. Compared to the AmC source which has the worst non-uniformity, the combined \Ge+Laser
source improves the $b$ term by 22.8\%.

\section{Optimization of Calibration Positions}
\label{sec:position}
\begin{figure}[!ht]
		\centering
		\includegraphics[width=0.48\textwidth]{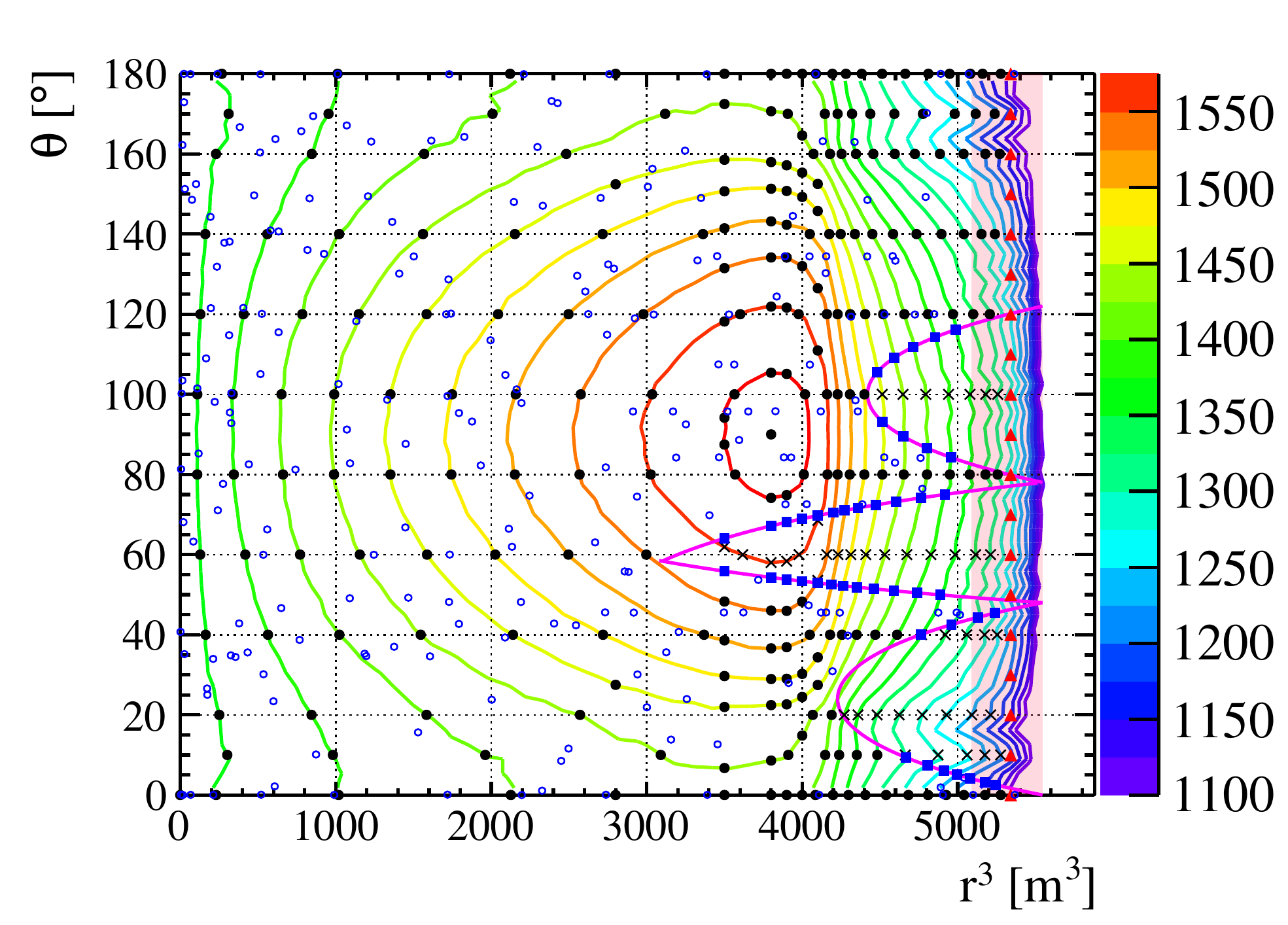}{\centering}		
		\caption{Contours of total number of PEs  on \thetaaa-\rrr ~plane and the arrangement
of calibration points for different cases. The three purple curves represent the CLS boundaries. 
The open circles represent the points from Ref.~\cite{junocollaboration2020calibration}. 
The black dots are common for Cases 2, 3 and 4. While the cross marks are for Case~2, the blue squares 
are for Case~3 and the red triangles are the positions from GT. }
		\label{fig:Positions}			
	\end{figure}

\noindent 
In Sec.~\ref{sec:source} we have looked into various calibration sources and found that 
\Ge+Laser combined source is the best choice for positrons in the kinetic energy range of [0-10]~MeV.
In addition to the source, the number and positions of calibration points should also be carefully
considered. 
As a simple measure of the detector energy response, the contours of total number of detected PEs 
on the $\theta-r^3$ plane for positron samples are drawn in Fig.~\ref{fig:Positions}. They clearly show that the 
detector energy response heavily depends on the position. Through finite calibration points, we try to capture 
the features of the detector energy response and then extrapolate to all areas as accurately as possible. The 
better we could do this, the more we will be able to improve the  energy uniformity of the detector.
However the calibration points from Ref.~\cite{junocollaboration2020calibration}, as represented by the open circles in Fig.~\ref{fig:Positions}, do not
have enough coverage near the detector boundary and the arrangement is somewhat random. We proposed to further optimize 
the calibration points utilizing the contours. The arrangement could be more efficient by assigning more (less) points 
in areas where the detector energy response changes dramatically (slowly).

For completeness, we considered a few different scenarios:
\begin{itemize}
	 \item  Case 1: mimics the ideal case with infinite calibration points. 2000 
points are randomly chosen in the ($r^5, \theta$) plane to allow more points in regions II and III
	\item Case 2: represents the optimal case with 275 points selected based on the contours of total number of PEs
	\item Case 3: corresponds to a more realistic case where those unreachable points in Case~2 are replaced with adjacent points on the
CLS boundaries
	\item Case 4: includes additional 19 points from GT on top of Case~3
\end{itemize}

\noindent The arrangement of the calibration points on the $\theta-r^3$ plane for the above cases~2-4 are 
also shown in Fig.~\ref{fig:Positions}. 
The three purple curves represent the CLS boundaries
and the areas they semi-enclose are not reachable. The points are the selected calibration positions.
The black dots are common for Cases 2, 3 and 4. While the cross marks are for Case~2, 
the blue squares are for Case~3 and the red triangles are the positions from GT.
Most of the selected points are at the intersections of the contours and fixed \thetaaa or \rrr 
~lines. In areas where the contours vary rapidly, more points are assigned. 

\begin{figure}[!ht]
		\centering
		\includegraphics[width=0.42\textwidth]{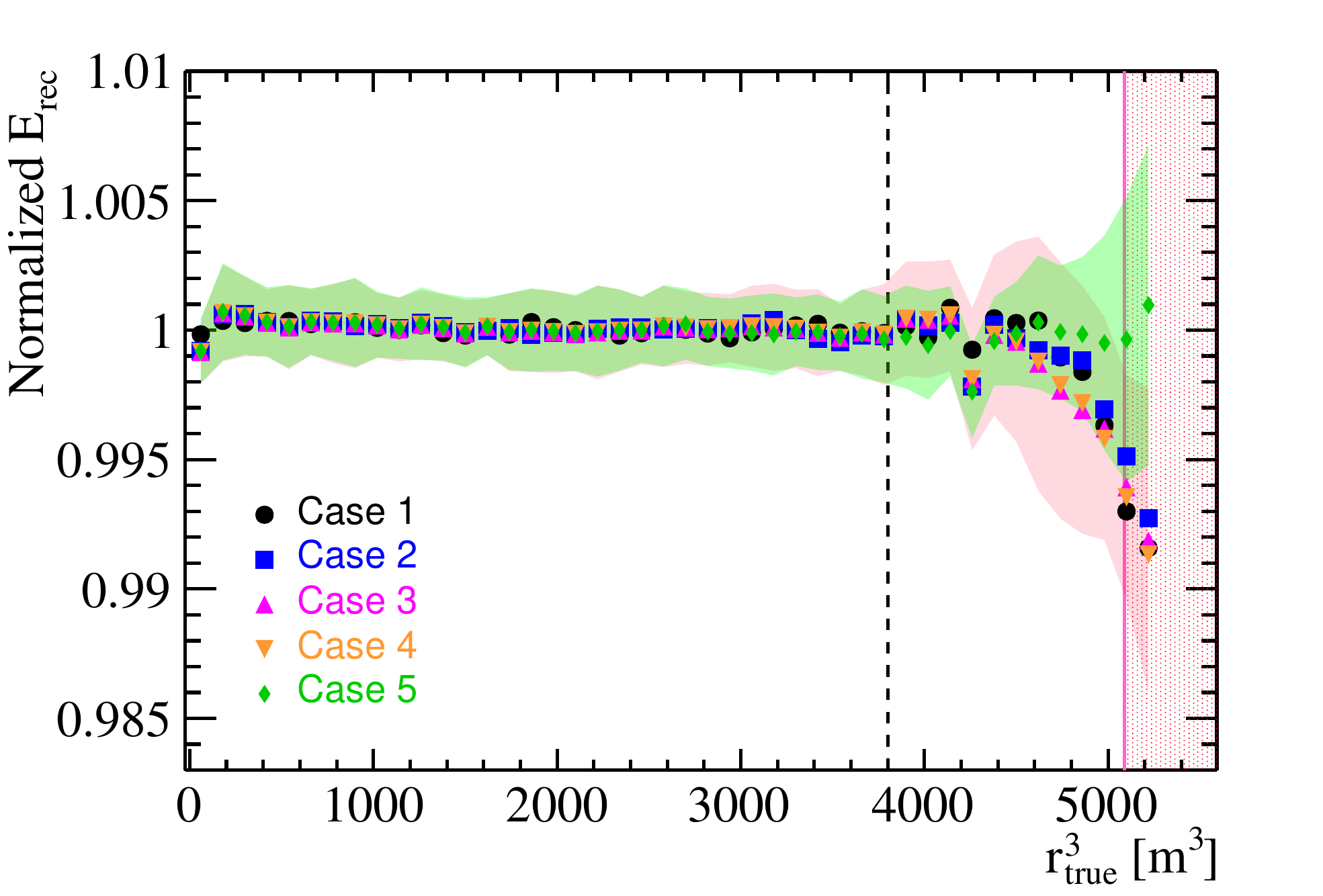}{\centering}		
		\caption{Uniformity of \Erec~with respect to \rrr from different cases of
calibration points at \Ek~= 5~MeV. The green and pink shadowed bands represent the 1$\sigma$ variation of the \Erec~along 
the $\theta$ direction for Cases 3 and 5 respectively. Case 5 will be discussed separately in Sec.~\ref{sec:phicorr}.}
		\label{fig:nonUniCompPositions}			
	\end{figure}

We chose the combined \Ge+Laser source, constructed the $\hat{\mu}$ maps for the 4 different cases 
above,  and then compared their energy reconstruction performance. Fig.~\ref{fig:nonUniCompPositions} 
shows the \Erec ~uniformity comparison at \Ek~= 5~MeV, similar results were observed at other energies 
given that energy uniformity has negligible energy dependence after using the combined source. 
Case~1 has the smallest non-uniformity. The difference between Case~1 and  
 Case~2 is marginal, indicating that we could largely reduce the total number of calibration 
points without jeopardizing the reconstruction performance. After replacing those unreachable points 
in Case~2, the energy uniformity becomes worse for Case~3 near the detector border. The GT system is
originally designed to calibrate the detector energy response near the detector border, complementary to 
CLS. After adding the points from the GT, the energy uniformity in Case~4 slightly improved with respect
to Case~3.
  
\begin{figure*}[!ht]
		\centering
		\includegraphics[width=0.95\textwidth]{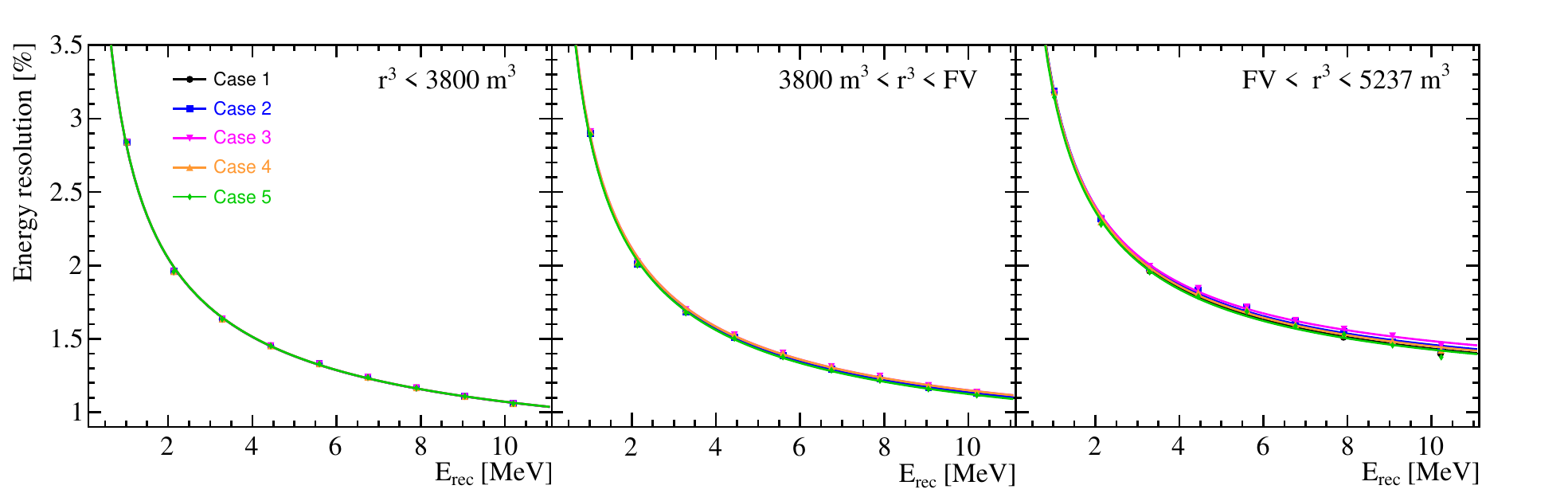}{\centering}		
		\caption{Comparison of the resolution of the reconstructed visible energy for positrons in the three regions of the CD. Different colors correspond to the various cases described in the text which explored the optimization of calibration positions, including azimuthal asymmetry for Case~5 discussed separately in Sec.~\ref{sec:phicorr}. The dots correspond to positrons with different kinetic energies. \Erec~is the mean value of the reconstructed visible energy. Please also note that energy non-linearity is not corrected here. }
		\label{fig:EresCompPositions}			
	\end{figure*}

The energy resolution was also compared for the four cases as shown in Fig.~\ref{fig:EresCompPositions}.
Overall, the performance is close. Differences of energy resolution ~in regions II and III are consistent with 
the energy uniformity comparison in Fig.~\ref{fig:nonUniCompPositions}, where smaller energy non-uniformity leads to better
energy resolution. The fitted energy resolution using Eqn.~\ref{eqn:EresFit} are listed in Tab.~\ref{tab:fitCompPositons}.
The $b$ term is larger for Case~3 compared to Case~2, and with the help of 
the GT, the $b$ term was slightly reduced from Case~3  to Case~4.

\begin{table}[h!]
  \begin{center}
    \caption{Comparison of fitted parameters for the empirical energy resolution for the five cases. The unit for a (b) is 
    \%$\times$MeV$^{\frac{1}{2}}$ (\%). The fit uncertainties for a and b are less than 0.005 (0.02) in regions I and II (III).}
    \label{tab:fitCompPositons}
    \begin{tabular}{ccccccc}
      \hline
      \hline
      \textbf{Region} & \multicolumn{2}{c}{\textbf{I}} & \multicolumn{2}{c}{\textbf{II}} & \multicolumn{2}{c}{\textbf{III}} \\
      \hline
         & a & b & a & b & a & b \\
                                
\hline
        Case 1 &                  2.76 & 0.622 & 2.80 & 0.711 & 3.02 & 1.07 \\
        Case 2 &                  2.76& 0.622 & 2.79 & 0.715 & 3.01 & 1.11 \\
        Case 3 &                  2.76 & 0.622 & 2.81 & 0.730 & 2.98 & 1.15 \\
        Case 4 &                  2.76 & 0.622 & 2.81 & 0.730 & 3.00 & 1.10 \\
        Case 5 &                  2.76 & 0.622 & 2.79 & 0.698 & 2.99 & 1.07 \\     
      \hline
      \hline
    \end{tabular}
  \end{center}
\end{table}


	\begin{figure*}[!ht]
		\centering
		\includegraphics[width=0.32\textwidth]{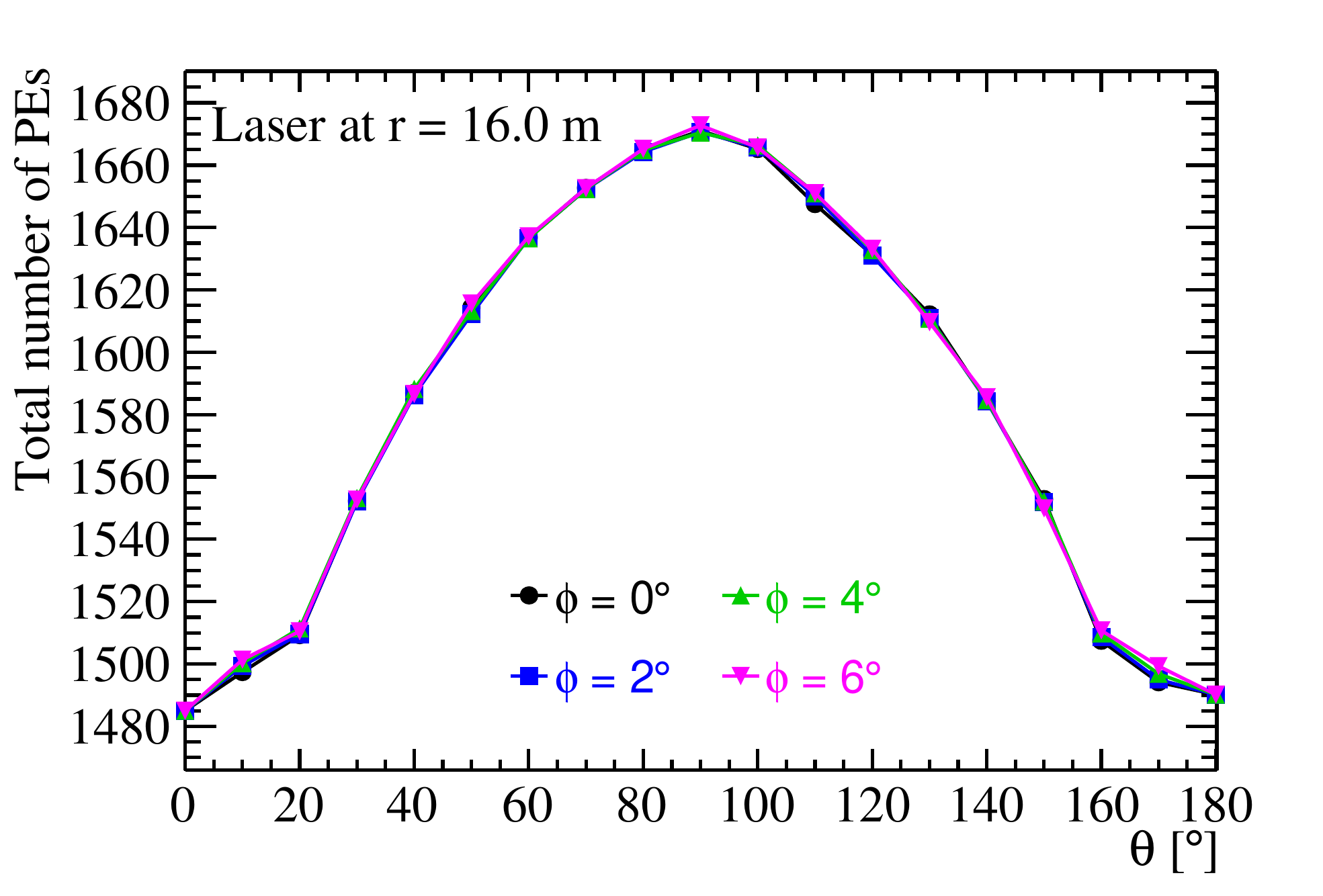}{\centering}
		\includegraphics[width=0.32\textwidth]{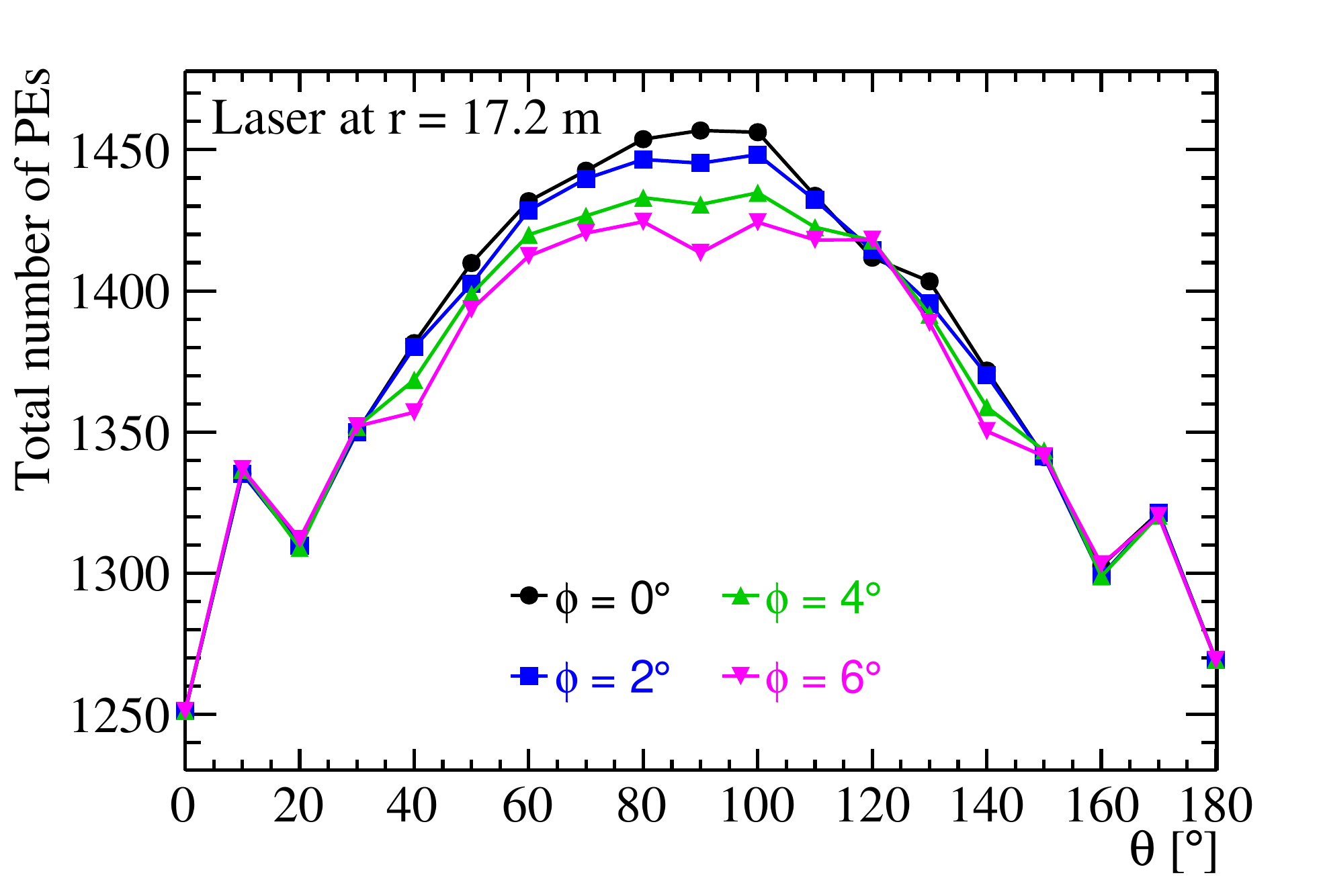}{\centering}
		\includegraphics[width=0.32\textwidth]{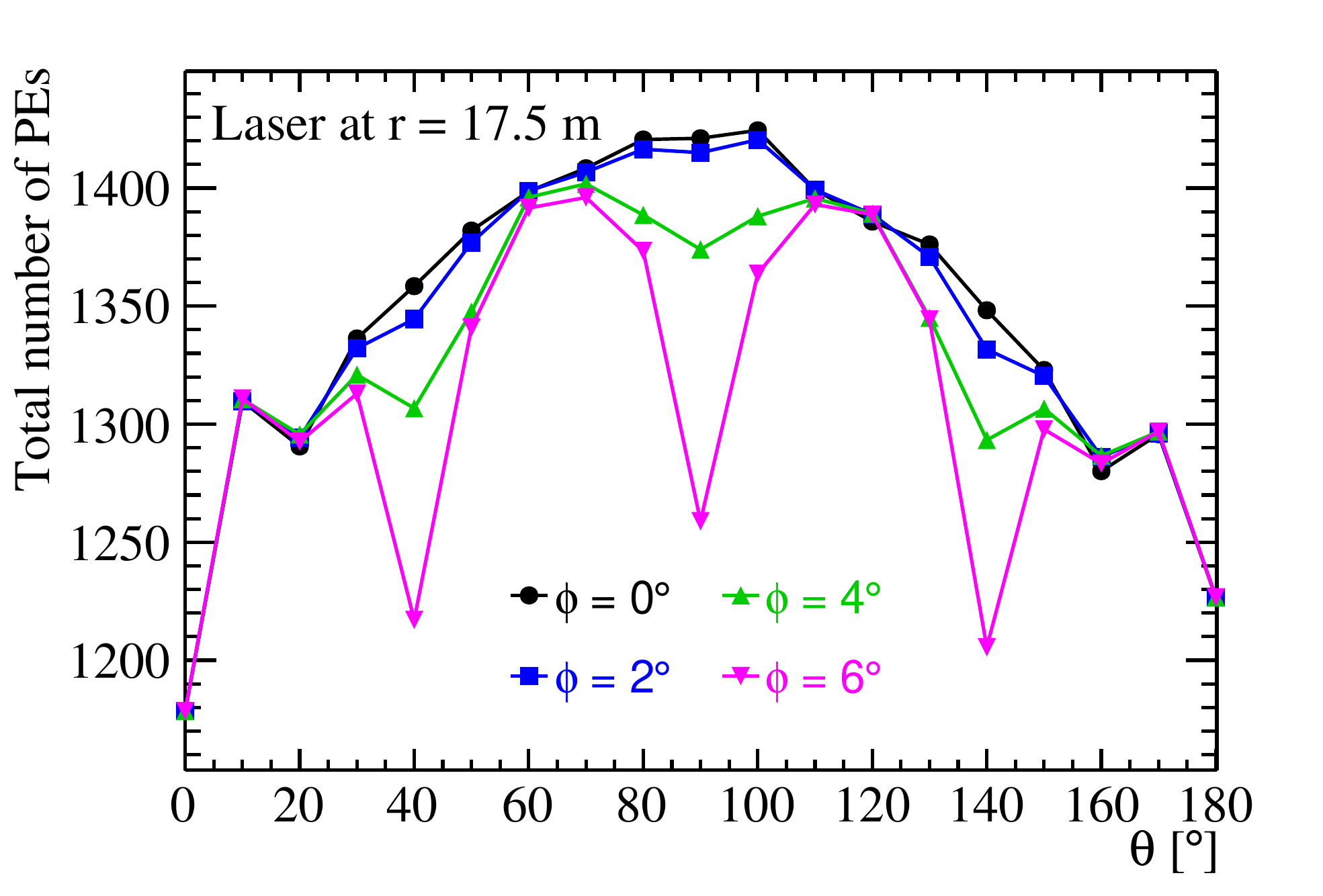}{\centering}
		\caption{Comparison of the distribution of total number of PEs as a function of \thetaaa among four different \phii planes.
From left to right, the radius is fixed at $r$ = 16~m, 17.2~m and 17.5~m respectively.}
		\label{fig:NPEphiComp}			
	\end{figure*}

\section{Residual Azimuthal Asymmetry}
\label{sec:phicorr}
\noindent After optimizing the calibration source and positions, there is still some residual energy non-uniformity
near the CD edge. 
While the $\hat{\mu}$ maps
have already taken into account the dependence on $r$ and $\theta$, the \phii dependence has not been considered
due to limitation of the calibration system. To check the \phii dependence, we compared the distribution of total number of PEs as a function of \thetaaa for two different \phii planes at three different radius. From 
the left plot to the right plot in Fig.~\ref{fig:NPEphiComp}, one can easily see that as the position gets closer
to the CD edge, the \phii dependence becomes more and more prominent.

The CD is only approximately symmetric along the \phii direction, since neither the PMTs nor the 
supporting structures have perfect azimuthal symmetry. Near the CD edge, the shadowing effect of the 
supporting structures can not be ignored any more. The dominant contribution comes from the numerous acrylic nodes.  
Although imperfect, the distribution of these acrylic nodes is roughly periodic every 6 degrees along 
the \phii direction. So we can apply a \phii dependent correction to the $\hat{\mu}$ maps within 6 degrees and extrapolate
to the full \phii range. Due to the mechanical limitation of CLS, we have to rely on MC simulation for 
this \phii correction. Three \phii planes with \phii= $2^\circ$, $4^\circ$, $6^\circ$ were selected, together with the CLS
plane where \phii= $0^\circ$, a correction function $f(\phi$) was produced and applied to the $\hat{\mu}$ maps of Case~3 from previous 
section for the edge region ($r > 15.6$~m) only. With this correction, the uniformity of \Erec ~at \Ek~= 5~MeV is plotted as
Case~5 in Fig.~\ref{fig:nonUniCompPositions}. 
 The improvement of the energy non-uniformity is outstanding in the edge region. After
the correction, the residual non-uniformity is about 0.17\% within the fiducial volume and 0.23\% across the entire detector.
 The energy resolution after the correction is plotted in Fig.~\ref{fig:EresCompPositions}, and the fitted results are listed in 
 Tab.~\ref{tab:fitCompPositons}, both referred to as Case~5. Comparing to Case~3, the improved energy uniformity propagates
to the energy resolution and leads to about 4.4\% and 7\% decrease for the $b$ term in regions II and III respectively.

The $f(\phi)$ correction derived above is able to reduce the residual energy non-uniformity in the CD edge region. One caveat is that 
this correction is derived from MC simulation, which has to be validated against real data. There are
several possible ways to do this. The calibration data from GT could be used to check this correction.
Another approach is to use spallation neutron events, which are abundant and uniformly distributed in 
the detector~\cite{dyb1230}. In the future, if we were able to reconstruct the event vertex with good precision, 
we could use spallation neutron events to obtain the $f(\phi)$  correction, instead of relying on MC simulation.

\section{Conclusion}	
\label{sec:summary}
\noindent It is rather challenging to achieve high precision energy resolution for large LS detectors such as JUNO.
Lots of studies have been done previously to address the energy uniformity utilizing calibration data in JUNO. In those studies, the residual 
energy non-uniformity is about 0.3\% within the fiducial volume, and gets much worse near the detector boundary due to complicated 
optical processes like total reflections, shadowing effects of opaque materials.
In this paper we expanded the $\hat{\mu}$ maps of expected PEs for PMTs to include the \thetaaa dependence. The choice of the 
calibration source was thoroughly investigated, and we found that \Ge+Laser combined source outperforms any single source 
across the entire energy range of interest. 
We also optimized the number and positions of
the calibration points based on a novel strategy, which utilizes the contours of the total number of detected PEs. 
The small residual non-uniformity caused by the azimuthal asymmetry of the detector was handled by a \phii dependent correction. 
As a result, we were able to reduce the energy non-uniformity from about 0.64\% to 0.38\% for regions near the detector boundary. And the energy non-uniformity within the fiducial volume (across the whole detector) could be well constrained under 0.17\% (0.23\%). As a direct consequence of the improved energy uniformity, better energy resolution was achieved.

Another interesting finding is that the energy non-uniformity in the central region of the detector is not particularly affected 
by the choice of the calibration source, nor by the asymmetry of the detector. This allows
for more flexibility on the calibration strategy in this  region. Moreover
 the detector energy response changes relatively slowly in this region so that only a small 
amount of calibration points are needed, which could also serve as a guideline for the calibration strategy.

In addition to the calibration sources, there will be various physics events occurring inside the LS detector as well, 
we should also be able to use them to obtain a better understanding of the detector response. 
Assuming we could select out some specific events, which are distributed across the entire detector, 
 and have reasonably well known energy and vertex, it would be straight forward to use them to construct
$\hat{\mu}$ maps  to include the \phii dependence as well. And to go one step further, if we could have 
huge amount of these events, we could try novel techniques such as Machine Learning to study the detector
energy response.
As we are entering the precision era of neutrino experiments, which demand much better energy resolution
than before, every bit of improvement counts. All the ideas and methods  in this paper
 improved the energy uniformity and consequently the energy resolution in JUNO. They  
 could be applied to other experiments with large LS detectors.
	
\section*{Acknowledgements}
\noindent This work was partially supported by the National Recruitment Program for Young Professionals, by the National Key R\&D Program of China under Grant No. 2018YFA0404100, by the Strategic Priority Research Program of the Chinese Academy of Sciences under Grant No. XDA10010100, and by the CAS Center for Excellence in Particle Physics (CCEPP).
We would like to thank Feiyang Zhang for the useful information and discussion about the 
calibration systems of JUNO.




\end{document}